\documentclass[letterpaper,english,letterpaper,english,prl,notitlepage,twocolumn]{revtex4-1}
\usepackage[T1]{fontenc}
\usepackage[latin9]{inputenc}
\usepackage{verbatim}
\usepackage{amsmath}
\usepackage{amssymb}
\usepackage{graphicx}
\usepackage{dsfont}

\makeatletter

\pdfpageheight\paperheight
\pdfpagewidth\paperwidth

\usepackage{color}
\usepackage[usenames,dvipsnames,svgnames,table]{xcolor}
\usepackage{graphicx}
\usepackage{comment}
\usepackage{ulem}

\definecolor{violet}{rgb}{0.58, 0.0, 0.83}

\newcommand{\gae}{\lower 2pt \hbox{$\,
\buildrel{\scriptstyle >}\over {\scriptstyle \sim}\,$}}
\newcommand{\lae}{\lower 2pt \hbox{$\,
\buildrel{\scriptstyle <}\over {\scriptstyle \sim}\,$}}

\newcommand{\fref}[1]{Fig.~\ref{#1}}
\newcommand{\eref}[1]{Eq.~\ref{#1}}

\makeatother

\newcommand{\addFN}[1]{{#1}}						

\usepackage{babel}
\begin{document}

\title{Topological Floquet-Thouless energy pump}


\author{Michael H. Kolodrubetz$^{1,2}$, Frederik Nathan$^{3}$, Snir Gazit$^{2}$, Takahiro Morimoto$^{2}$, Joel E. Moore$^{1,2}$} 
\affiliation{$^1$Materials Sciences Division, Lawrence Berkeley National Laboratory, Berkeley, CA, USA
  \\$^2$Department of Physics, University of California, Berkeley, CA, USA
  \\$^3$Center for Quantum Devices, Niels Bohr Institute, University of Copenhagen, Copenhagen, Denmark} 

\begin{abstract}
  We explore adiabatic pumping in the presence of periodic drive, finding a new phase in which the topologically quantized pumped quantity is energy rather than charge. The topological invariant is given by the winding number of the micromotion with respect to time within each cycle, momentum, and adiabatic tuning parameter. We show numerically that this pump is highly robust against both disorder and interactions, breaking down at large values of either in a manner identical to the Thouless charge pump. Finally, we suggest experimental protocols for measuring this phenomenon.
\end{abstract}
\maketitle

The Thouless charge pump serves as a simple yet fundamental example of topology in quantum systems \cite{Thouless1983_1}. The hallmark of this effect is the transport of a precisely quantized amount of charge during an adiabatic cycle in parameter space. This remarkable phenomenon has been demonstrated experimentally in various physical systems such as few-body semiconductor quantum dots \cite{SwitkesPump,BlumenthalPump,Buitelaar2008,Giazotto2011}
and more recently in a one-dimensional chains of ultra-cold atoms trapped in an optical lattice \cite{Lu2016,Nakajima2016,Lohse2016}.

Recently, the classification of topological phases of matter has been extended
to periodically driven (Floquet) systems far from equilibrium \cite{Kitagawa2010,Jiang2011,Rudner2013_1,Nathan2015,Roy2016}.
In particular, periodic driving can lead to new topological phases that have no analogy in undriven systems
~\cite{Ho2012,Khemani2016,Roy2016_2,Titum2016,Keyserlingk2016,Else2016,Potter2016,Keyserlingk2016_2,Roy2016_2,Zhou2016,Po2016,Harper2017,Potter2017,Roy2017,PoArxiv2017,PotterArxiv2017}, an idea which has been confirmed experimentally \cite{Hu2015,Mukherjee2017}. A natural question to ask is whether these far-from equilibrium systems can also exhibit new topological pumping effects?

\begin{figure}[t!]
\includegraphics[width=0.7\columnwidth]{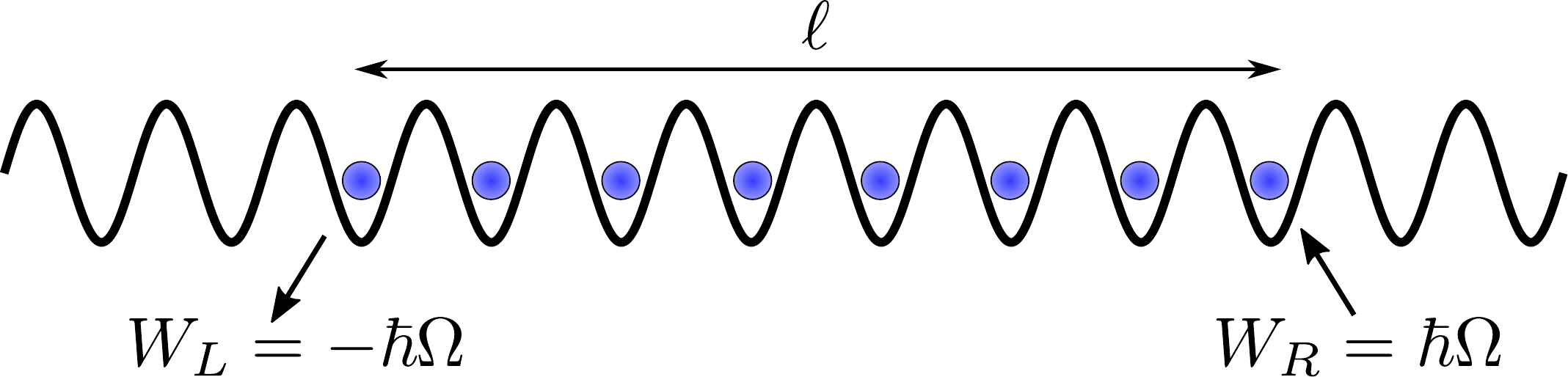}
\caption{\label{fig:intro} Illustration of the topological energy pump. Upon ramping the pump adiabatically around a cycle, the filled region of length $\ell \gg 1$ remains localized, but nevertheless quantized work is performed at the edges of the filled region in quanta of the drive energy $\hbar \Omega$.}
\end{figure}

In this paper, we  answer this question in the affirmative by explicitly constructing a generalized adiabatic pump in a Floquet system. We find a novel phase in which energy, rather than charge, undergoes quantized pumping. Specifically,  upon adiabatic cycling of a particular parameter, partially filled systems in this phase transport  energy from one side of the filled region to the other, as illustrated in \fref{fig:intro}.
The energy transported per cycle is quantized in units of the drive frequency $\hbar\Omega$.

Using numerical and analytical arguments, we show that this phenomenon is stabilized by disorder and, via many-body localization, remains robust in the presence of interactions. In this way, we demonstrate the existence of a new stable topological pump that can only be realized in the presence of periodic driving.

\begin{figure}
\includegraphics[width=\columnwidth]{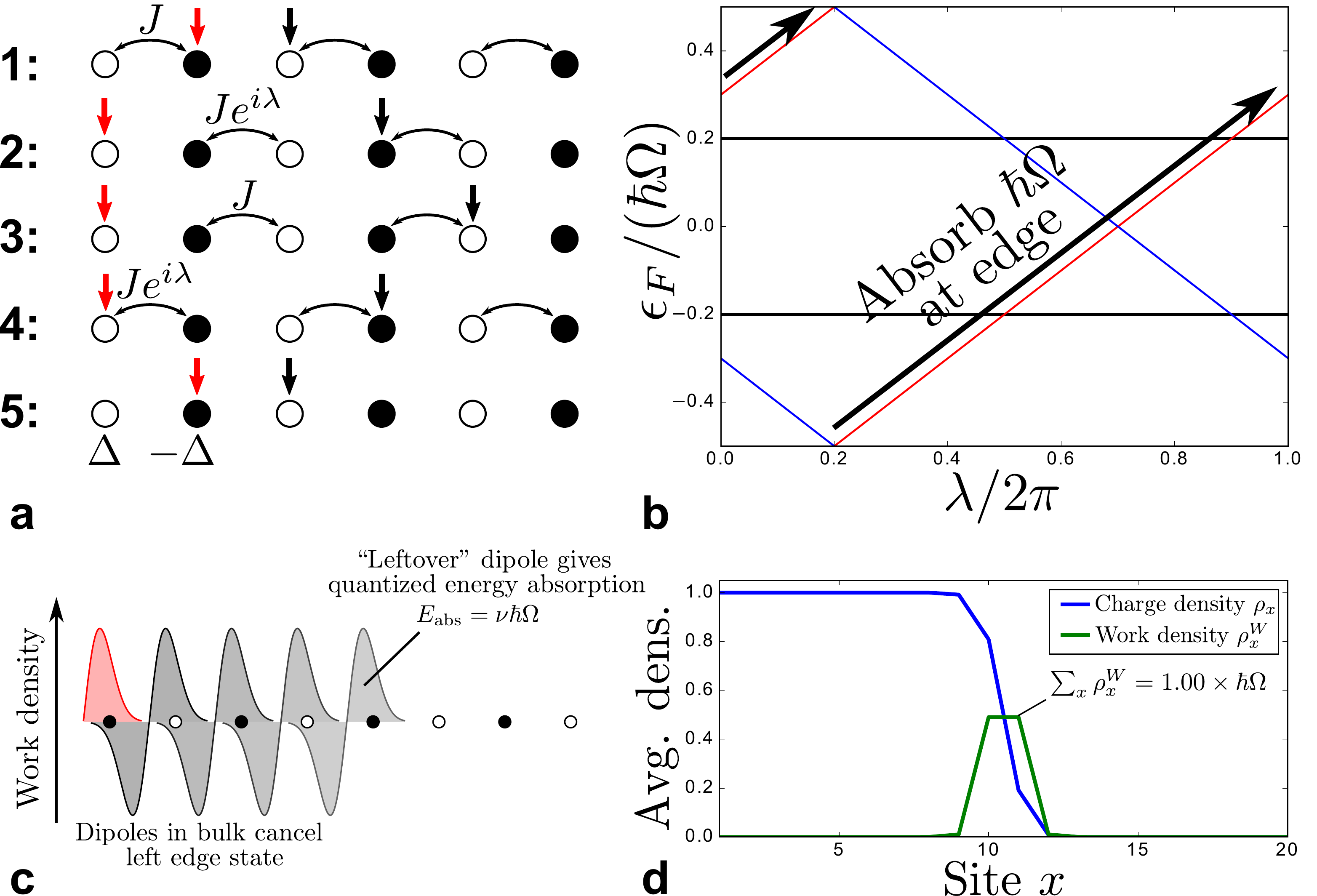}
\caption{\label{fig:setup_and_models} (a) Illustration of the anomalous Floquet pump (\eref{eq:H_flat_band}), which involves five steps of period $T/5$ with fined-tuned hopping $J_\mathrm{tuned} = 5 \hbar \Omega / 4$. Red and black arrows trace the positions of edge and bulk states respectively. (b) Quasienergy spectrum as a function of the tuning parameter $\lambda$ show bulk bands (black), left edge state (red), and right edge state (blue).
(c) Illustration of response measured in numerics, for which only the left half of the system is filled. (d) Numerical results for the local work and charge density for the model in \eref{eq:H_flat_band} averaged over a single ramp from $\lambda=0$ to $2\pi$ with $L=20$, $N_c=12$, and $N_\lambda=1$. Data for $\rho^W_x$ is in units of $\hbar \Omega$.}
\end{figure}

\emph{Model.}
Let us begin by introducing a simple model that exhibits topological energy pumping, which we will later demonstrate is topologically robust to perturbations.
The model consists of a five-step driving protocol, with Hamiltonians $H_j = h_j+h.c.$, where
\begin{align}
\nonumber
&h_1=-J\sum_{x} c^{\dagger}_{A,x}c_{B,x} , \quad \hspace{3.7mm} h_2=-J\sum_{x} e^{i \lambda} c_{B,x}^{\dagger}c_{A,x+1} 
\\
\nonumber
&h_3=-J\sum_{x} c_{B,x}^{\dagger}c_{A,x+1} ,\quad h_4=-J\sum_{x} e^{i \lambda} c_{A,x}^{\dagger}c_{B,x} ,
\\&h_5=\frac{\Delta}{2}\sum_{x} \left(c_{A,x}^{\dagger}c_{A,x} - c_{B,x}^{\dagger}c_{B,x}\right)
  \label{eq:H_flat_band}
\end{align}
acting on $L$ sites with open boundary conditions. The protocol is chosen to be time periodic with $H(t)=H(t+T)$ such that $H(0 < t < T/5)=H_1$, $H(T/5 < t < 2 T /5)=H_2$, etc. This model is particularly simple if the tunneling strength $J$ takes the value $J_\mathrm{tuned} \equiv 5 \hbar \Omega /4$, where $\Omega = 2\pi / T$. At this fine-tuned point, the fermions hop exactly one site at each step, such that a fermion initialized at any site returns to the same site after one driving cycle, as illustrated in \fref{fig:setup_and_models}a.

Using the Floquet formalism, we write the single-particle time evolution $U$ in the form $U(t)=\mathcal P(t)e^{-iH_F t}$, where the micromotion $\mathcal P(t)=\mathcal P(t+T)$ describes the dynamics within each cycle and $H_F$ is the effective Hamiltonian that describes stroboscopic behavior at multiples of the period $T$ \cite{Bukov2015_2}. For $J=J_\mathrm{tuned}$, the Floquet eigenstates are localized states $|x,\alpha\rangle \equiv c_{\alpha,x}^\dagger |\mathrm{vac}\rangle$. The eigenvalues of $H_F$, known as quasienergies, are only well-defined modulo $\hbar \Omega$. For a particle initially located on a site in the bulk, the phase $e^{i\lambda}$  acquired during step $2$ is cancelled by the phase $e^{-i\lambda}$ during step $4$, yielding flat \addFN{quasienergy} bands at $\epsilon^F_\mathrm{bulk} = \pm \Delta / 5$. However, a particle initially located at site $|1,B\rangle$ or $|L,A\rangle$ is  unable to hop during steps 2 and 3, causing it pick up a $\lambda$-dependent phase during the driving cycle, which translates into a $\lambda$-dependence of these edge state quasienergies (\fref{fig:setup_and_models}b). While the bulk bands are trivial and can be shown to have vanishing Chern number with respect to $\lambda$ and quasimomentum $k$ \footnote{Explicitly, if  $|u_{F}^{\alpha}(k,\lambda){\rangle}$ is a single-particle Floquet eigenstate in band $\alpha$ with quasienergy $\epsilon_F^\alpha(k,\lambda)$, then the Floquet Chern number  $C_1^{F,\alpha} \equiv \frac{i}{2\pi} \int d\lambda dk \left({\langle} \partial_k u_F^\alpha | \partial_\lambda u_F^\alpha {\rangle} - h.c. \right)$ vanishes. Note that $C_1^F$ is independent of the choice of origin $t_0$ used to define $H_F = i \log(U[t_0 \to t_0 + T)]/T$.},
the edge states (red and blue) clearly exhibit topologically nontrivial winding. The question, then, is how to characterize and measure the topological properties of this model?

\emph{Topology and measurement.} The main insight for understanding our model comes from noting that the band structure in \fref{fig:setup_and_models}b is identical to that found in the two-dimensional anomalous Floquet insulator (cf.~Fig.~1 in Ref.~\cite{Rudner2013_1}) with the the pump parameter $\lambda$ playing the role of momentum $k_y$. 
In this way, our model is a dimensionally reduced version of the anomalous Floquet insulator~\cite{Rudner2013_1,Titum2016}, in the same way that the Thouless pump may be thought of as the dimensional reduction of a Chern insulator. This immediately implies the existence of a topological invariant characterizing our pump, namely the winding number of the micromotion,
\begin{equation}
  \nu=\frac{1}{8\pi^2} \int dt d\lambda dk \mathrm{Tr} \left(\left[\mathcal{P}^\dagger \partial_\lambda \mathcal{P}, \mathcal{P}^\dagger \partial_t \mathcal{P} \right] \mathcal{P}^\dagger \partial_k \mathcal{P}\right),
  \label{eq:winding_number}
\end{equation}
defined on the compact three-dimensional parameter space $(t,\lambda,k)$. While the micromotion and thus the winding number in principle depend on the branch cut defining $H_F$, the fact that Chern numbers of the bulk bands vanish implies that the winding number is independent of this choice~\cite{Rudner2013_1}. For the model we consider here, $\nu=1$.

One hint for the observable consequences of this topological index comes from examining the quasienergy spectrum in the presence of open boundary conditions (\fref{fig:setup_and_models}b). Upon adiabatically ramping $\lambda$ from $0$ to $2\pi$, the bulk remains unchanged while the left (right) edge state wraps around the Floquet Brillouin zone, absorbing (emitting) a quantum of energy. Upon completing the cycle, the system returns to its initial electronic state. Therefore the nontrivial topology does not lead to any direct pumping of the fermions. Instead, as we will show, ramping $\lambda$ performs quantized work on the external driving fields. 

Specifically, we now show that the quantized observable is the $\lambda$-averaged ``force polarization'' $P_F \equiv \sum_x x \rho^F_x$, where
\begin{equation}
  \rho^F_x = \frac{1}{2}\left \langle \left\{ \sum_\alpha c^\dagger_{\alpha,x\alpha} c_{\alpha,x} ,\partial_\lambda H \right\} \right \rangle
  \label{eq:rho_F}
\end{equation}
is the local generalized force required to change $\lambda$ by a small amount and $\alpha=\{A,B\}$ sums over sublattices \footnote{One may readily see this by analogy: if $H(x)$ is a complicated potential acting on a point particle due to the fermions in the lattice, then $-{\langle} \psi | \partial_x H | \psi {\rangle}$ is the force acting on $x$. Note that this is true for arbitrary state $\psi$, whether or not in equilibrium.}.
Changing $\lambda$ by a finite amount thus requires a local work
\begin{equation*}
  \rho^W_x = \int \rho^F_x\left[\lambda(t),t\right] \dot \lambda(t) dt.
\end{equation*}
While this expression holds for arbitrary non-equilibrium systems, it becomes independent of speed in the limit of slow ramps. Thus a finite work polarization $P_W = \int P_F d\lambda$ implies that work is done on one half of the system and done by the other half. We will see that quantization of $P_W$ thus implies that this differential work is quantized, as illustrated in \fref{fig:intro}.

Quantization of $P_W$ follows immediately from dimensionally reducing the anomalous Floquet insulator, as the topologically quantized magnetization \cite{NathanArxiv2016} immediately reduces to $P_W$. In practice, the work polarization may be directly measured by filling part of the system and measuring the time-dependence local force $\rho^F_x$ near the edges of the filled region, as illustrated in \fref{fig:intro}. Within the fully filled or fully empty regions nothing is able to move, hence no work is done: $\rho^W_x = 0$. Furthermore, as the net work on the entire system vanishes, the work done near the left edge of the filled region, $W_L$, must exactly cancel that done near the right edge: $W_R=-W_L$. For a filled region of length $\ell$ lattice sites which is much larger than the localization length $\xi$, the work polarization per particle is then given by $P_W^\mathrm{tot} \approx (W_R - W_L) \ell / 2$. As the average work polarization per filled unit cell is quantized to be $\overline{P}_W=\nu \hbar \Omega$, we also have $P_W^\mathrm{tot} = \nu \hbar \Omega \ell$. Equating these expressions, we find that
\begin{equation}
  W_R=-W_L=\nu \hbar \Omega.
\end{equation}
Further details on this derivation may be found in the supplement \cite{Supplement_Energy_Pump_Short}.

To confirm these predictions, we consider a slightly different setup in which we fill only the left half of the system, i.e., sites $1$ through $L/2$. Then the only contribution to the force comes from the density step at $L/2$, such that the entire system absorbs/emits an integer number of photon quanta. \fref{fig:setup_and_models}c illustrates how this emerges from adding the quantized polarization in each localized state. Numerically, we start from this initial state and ramp $\lambda$ from $0$ to $2\pi N_\lambda$ at a constant rate $\dot \lambda = 2 \pi / (N_c T)$. While slow time-dependence of $\lambda$ formally breaks the $T$-periodicity, it has been shown than an appropriate extension of adiabaticity may be defined \cite{Hone1997_1,Drese1999_1,Weinberg2017}, which is nevertheless subtle due to the presence of resonances which must be crossed diabatically. In practice, we find that an appropriate adiabatic limit is reached for $N_c \gg 1$ and ramping over many adiabatic cycles ($N_\lambda \gg 1$) to remove initial transients. We then expect the total energy absorbed by the system, 
\begin{equation}
  E_\mathrm{abs} \equiv \int \langle \partial_\lambda H \rangle \dot \lambda dt,
\end{equation}
to be quantized in units of $\hbar \Omega$. In the supplement \cite{Supplement_Energy_Pump_Short} we show this analytically for our simple model, and we verify this numerically in \fref{fig:setup_and_models}d.

\emph{Disorder and interactions.}
 Having determined the basic properties of our topological energy pump in an analytically tractable limit, we now demonstrate its robustness to disorder and interactions. One might naively expect this robustness to be trivial, as topological states are often argued to be protected against weak perturbations. However, in the presence of disorder, the ability to adiabatically track a given localized eigenstate is known to be ill-defined, as the eigenstate will undergo weakly avoided crossings on arbitrary
length scales \cite{Khemani2015}. We will address this issue analytically in a follow up work \cite{Upcoming_Long_Paper}, but for now we provide numerical support regarding its stability.

Specifically, we add static chemical potential  disorder to our Floquet system,
\begin{equation}
  H_\mathrm{dis}=\sum_{\alpha,x} w_{\alpha,x} c_{\alpha,x}^\dagger c_{\alpha,x}~,
\end{equation}
where the disorder is drawn from a box distribution $w_{\alpha,x}/\Omega \in [-W,W]$. We also consider deviating from the fine-tuned limit by a ``detuning'' $\alpha$ \footnote{This choice of detuning $\Delta$ and $J$ simultaneously is not unique. Other choices will give similar results.}.
such that 
\begin{equation}
  \Delta=\alpha \Omega~,~J=J_\mathrm{tuned} (1-\alpha).
  \label{eq:alpha_detuning}
\end{equation}
We then carry out the same procedure as in \fref{fig:setup_and_models}c to measure topological energy absorption.

\begin{figure}
\includegraphics[width=1\columnwidth]{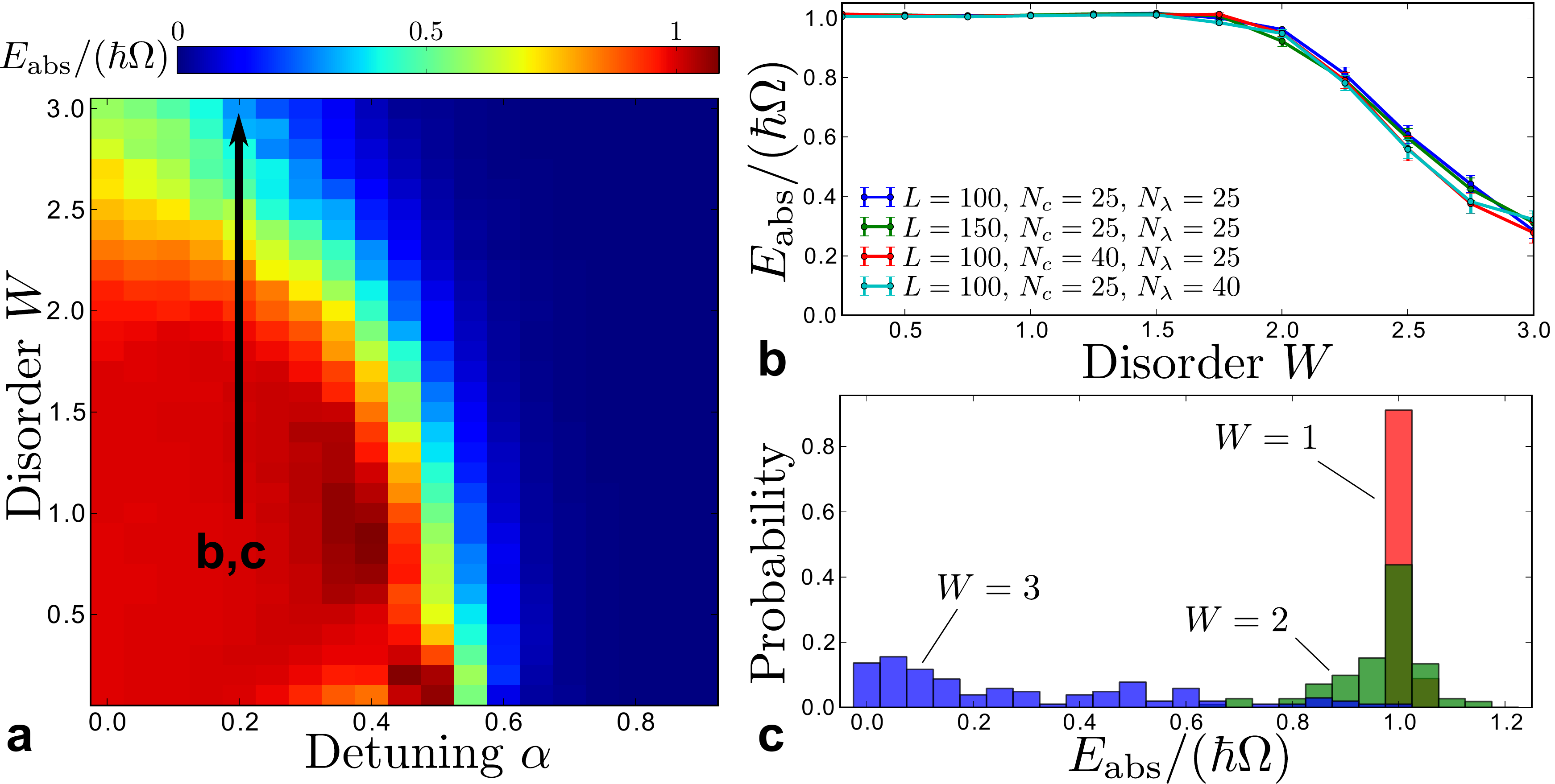}
\caption{\label{fig:anderson_localization} (a) Phase diagram of energy pump as a function of disorder $W$ and detuning $\alpha$ (see \eref{eq:alpha_detuning}) at fixed $L=100$ and $N_c=N_\lambda=25$. In the absence of disorder a phase transition occurs at $\alpha=1/2$. In the presence of disorder, the topological plateau appears stable over a wide region. (b) Cut at fixed $\alpha=0.2$ showing a slow crossover to topologically trivial independent of $L$, $N_c$, and $N_\lambda$. (c) Histogram over disorder configurations of $E_\mathrm{abs}$ at three points along the crossover showing the breakdown of topological quantization.}
\end{figure}

The disorder-averaged phase diagram for a wide range of disorder strengths and detunings is shown in \fref{fig:anderson_localization}a. There is clearly a wide region with well-quantized energy pumping (red), up to disorder strengths and detuning of order $\hbar \Omega$. In fact, for the majority of the phase diagram, disorder is actually necessary to see quantization of the energy transport. The simplest reason for this is that, in the absence of disorder, any generic model will not be localized and our measurement of $E_\mathrm{abs}$ at the localized density edge is not meaningful. This is seen in our phase diagram for $\alpha \neq 0$, where a small amount of disorder clearly improves the quantization for the system size shown. Furthermore, we will show in a follow up work \cite{Upcoming_Long_Paper} that even the appropriately defined clean limit of $P_F$ has a non-topological contribution which is suppressed by localization. In either case, the phase diagram clearly shows a large nearly quantized plateau at weak disorder below the topological transition at $\alpha=1/2$. For instance, the data in \fref{fig:anderson_localization}b is quantized to within $0.4\%$ and $0.8\%$ at $W=1$ and $3 / 2$ respectively for $L=150$, $N_c=N_\lambda=40$.

At large disorder strengths, we expect a topological transition to a trivial state while maintaining Anderson localization throughout \footnote{Anderson localization should always exist, even in the presence of driving, for this one-dimensional model \cite{Anderson1958,Mott1961,Agarwal2017}.}.
Surprisingly, we instead find a slow crossover for which energy is still pumped, but not quantized. This is unlike the sharp transition found in the anomalous Floquet Anderson insulator \cite{Titum2016}, and illustrates a fundamental difference regarding the role of disorder in one dimensional pumps compared to their higher-dimensional counterparts. 
For the energy pump, one of the tuning parameters, $\lambda$, couples strongly to the quasienergies, even when the system is localized. For the anomalous Floquet Anderson insulator,  the winding number is defined as in Eq.~\eqref{eq:winding_number} with angles $\theta_x$ and $\theta_y$ defining twisted boundary conditions in place of the parameters $\lambda$ and $k$. For that model, the localization of  Floquet eigenstates implies that the change of quasienergy due to either twist angles is exponentially suppressed. In contrast, the ``dimensional extension''  of the energy pump features Floquet states that are delocalized in the $y$-direction. Hence the quasienergy spectrum is sensitive to changes of $\theta_y$, i.e., $\lambda$.


The breakdown of topological energy pumping may be traced to this increased sensitivity to $\lambda$. As the disorder strength $W$ is increased, the $L$ individual quasienergy mini-bands $\varepsilon_n (\theta,\lambda)$ may undergo topological gap closings and reopenings, potentially introducing non-trivial Chern numbers. This yields a Floquet branch cut dependence of the winding number $\nu(\epsilon_\mathrm{gap}^F)$ \cite{Rudner2013_1}. As our measurement populates quasienergy states at random (the ``infinite temperature'' ensemble), we stochastically sample over these winding numbers. Thus the non-quantized energy pump may be thought of as an average of the topological winding number over both gaps and disorder realizations. 


This argument is consistent with the histogram of $E_\mathrm{abs}$ in this crossover region (\fref{fig:anderson_localization}c), which shows broadening from a perfectly quantized $\delta$-function peak at $E_\mathrm{abs} = \hbar \Omega$ towards statistical ensemble that will eventually be non-topological ($\overline{E}_\mathrm{abs} = 0 $). Importantly, this breakdown by a proliferation of Berry monopoles is precisely the  mechanism that leads to the loss of charge pump quantization in disordered systems \cite{Onoda2006,Chern2007}. Thus the crossover behavior in our system likely falls into the same class as this undriven case.

\begin{figure}
\includegraphics[width=1\columnwidth]{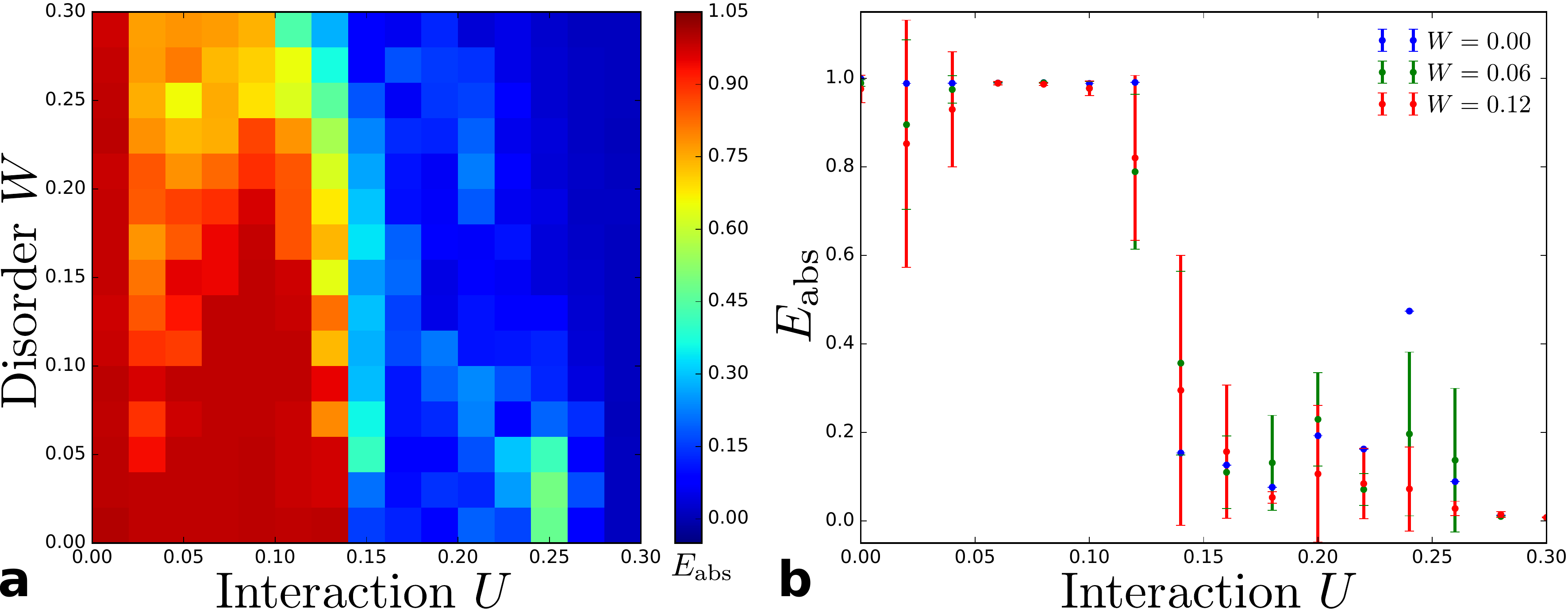}
\caption{\label{fig:many_body_localization} Phase diagram of energy pumping in an interacting many-body localized (MBL) system for $L=16$, $N_\lambda=64$, and $N_c=256$. Error bars in (b) show standard error over a fixed number of disorder configurations. Interestingly, the error bars become smaller -- indicating increased stability of the MBL energy pump -- in the presence of weak nonzero interactions. $E_\mathrm{abs}$ and $U$ are in units of $\hbar \Omega$.}
\end{figure}

\emph{Many-body localization.} Finally, let us see that our results hold in the presence of many-body localization. We test this by adding nearest neighbor interactions 
\begin{equation}
  H_\mathrm{int} = U\sum_{j} \left(n_j-\frac 1 2\right) \left(n_{j+1}-\frac{1}{2}\right)
\end{equation}
throughout the cycle and simulate the dynamics via exact diagonalization \footnote{Here we mean nearest neighbors independent of sublattice, i.e., $|1B{\rangle}$ neighbors $|1A{\rangle}$ and $|2A{\rangle}$.}.
In \fref{fig:many_body_localization}a, we map out the phase diagram as a function of interaction and disorder strengths. The data confirm  that the energy absorption remains beautifully quantized in the topological phase (\fref{fig:many_body_localization}b). Interestingly, the data  indicate that weak interactions also stabilize the topological phase. While this may be due to a trivial microscopic effect such as shortening of the localization length due to interactions, it leaves open the tantalizing possibility that interactions stabilize the phase and lead to an energy pump that is again topologically protected.

\emph{Experiments.} The topological energy pump is directly amenable to being realized experimentally, requiring hopping models in one dimension similar to those recently realized in optical lattice charge pumps \cite{Lu2016,Nakajima2016,Lohse2016}. Instead of measuring local charge, these experiments would simply have to measure local force, $\rho^F_x$. This should be readily realized by combining adiabatic pump protocols with systems that enable site-resolved measurement, such as optical lattice microscopes \cite{Bakr2009,Cheuk2015}, trapped ion arrays \cite{Blatt2012_1}, and other engineered platforms \cite{Barends2013_1,Roushan2014_1,BernienArxiv2017}. In addition to the pulsed multi-step protocols discussed in this work, which are quite natural in such engineered systems, we will show elsewhere that 
the topological pumping may also occur in 
  monochromatically driven models, such as a driven version of the Rice-Mele model \cite{Rice1982,Upcoming_Long_Paper}. This opens the intriguing possibility to directly measure the back-action on the drive lasers. For instance, if the periodic  driving is realized by a pair of Raman lasers with frequency difference $\Omega$, adiabatic cycling of the pump parameter $\lambda$ would result in quantized transfer of $\nu$ photons from one Raman beam to the other. If one further quantizes the Floquet drive photons, for instance by use of a high-Q cavity, then each adiabatic cycle would directly back-act on the cavity photons. This can, for example, lead to either quantized absorption/emission of cavity photons, whose behavior at low photon number represents an interesting quantum limit of our problem.

\emph{Conclusion.}
We have introduced a novel topological energy pump which exhibits a new type of topologically protected response with no equivalent in undriven systems. The pump is inspired by a dimensional reduction scheme from the anomalous Floquet insulator, but features  fundamentally different topological protection and transport properties. We note that other topological energy pumps recently introduced in the driven qubit systems derive instead from reducing the Thouless charge pump to zero dimensions, replacing momentum with a magnetic field angle \cite{Weinberg2017} or the phase of a second incommensurate drive \cite{MartinArxiv2016}. This suggests a number of fascinating future directions from dimensional reduction of other entries in the Floquet periodic table \cite{Nathan2015,Roy2016}, such as the Floquet generalization of the $\mathbb{Z}_2$ pump \cite{Fu2006, Essin2007}	 or fractionalized systems \cite{Grushin2014_1}. Furthermore, studying the back-action of our topological pump on a classical or a quantum drive represents an interesting quantum adiabatic limit on statistical mechanics, where pumping of bosonic objects such as the drive photons is a long sought-after goal \cite{Chamon2011,Ludovico2016}.

\emph{Acknowledgments.} We acknowledge useful discussions with C. Chamon, C. Laumann, N. Nagaosa, A. Polkovnikov, M. Rudner, and D. Stamper-Kurn. MHK and JEM were supported by Laboratory directed Research and Development (LDRD) funding from Berkeley Laboratory (LBNL), provided by the Director, Office of Science, of the U.S. Department of Energy (DOE) under Contract No. DEAC02-05CH11231, and from the U.S. DOE, Office of Science, Basic Energy Sciences (BES) as part of the TIMES initiative. JEM and SG acknowledge additional support from the Simons foundation. SG was supported by the National Science Foundation (NSF) under grant number DMR-1411343. TM was supported by the Moore Foundation and the Quantum Materials program at LBNL. FN is grateful to the Villum Foundation and the Danish National Research Foundation for support. This work was partially performed at the Aspen Center for Physics (NSF grant PHY-1607611) and the Kavli Institute for Theoretical Physics (NSF grant PHY-1125915). Computational work was done on the Lawrencium cluster at LBNL.

\bibliography{floquet_energy_pump_references}

\appendix

\begin{widetext}

\section{Dimensional reduction and quantized response}

Let us begin by showing that the dimensional reduction of magnetization
gives quantization of the work polarization. In Ref.~\cite{NathanArxiv2016}, it is shown
that the time-averaged magnetization density $\left\langle M\right\rangle=\sum_{n=1}^{N}\left\langle M\right\rangle _{n}$
of the anomalous Floquet Anderson insulator after summing over localized
single particle Floquet eigenstates $|\psi_{n}\rangle$ is quantized
as 
\[
\left\langle M\right\rangle =\frac{N}{T}\nu,
\]
where $N=N_{\mathrm{orb}}L^{2}$ is the number of states for an $L\times L$
system with $N_{\mathrm{orb}}$ orbitals/sublattices and both the
particle charge $q$ and lattice constant $a$ are set to $1$. This
may be rewritten as average quantization of $\left\langle M\right\rangle _{n}$:
\begin{eqnarray*}
\overline{\left\langle M\right\rangle _{n}}\equiv\frac{\left\langle M\right\rangle }{N}=\frac{\nu}{T} & = & \frac{1}{N}\sum_{n}\left(\frac{1}{2T}\int_{0}^{T}dt\langle\psi_{n}|{\bf r}\times\partial_{t}{\bf r}|\psi_{n}\rangle\right)\\
 & = & \frac{1}{N}\sum_{n}\left(\frac{1}{2T}\int_{0}^{T}dt\langle\psi_{n}|\left[\hat{x}\partial_{t}\hat{y}-\hat{y}\partial_{t}\hat{x}\right]|\psi_{n}\rangle\right).
\end{eqnarray*}
Noting that the velocity operator is $\partial_{t}\hat{y}=(L/2\pi\hbar)\int dk_{y}(\partial\hat{H}/\partial k_{y})$
and utilizing antisymmetry of the integrand with respect to $k_{x}$
and $k_{y}$, we see that
\begin{equation}
\frac{\nu}{T}=\frac{1}{2\pi\hbar N_{\mathrm{orb}}L}\left[\int dk_{y}\sum_{n}\left(\frac{1}{T}\int_{0}^{T}dt\langle\psi_{n}|\hat{x}\partial_{k_{y}}H|\psi_{n}\rangle\right)\right].\label{eq:magn_quant_per}
\end{equation}
Upon dimensional reduction, $k_{y}\to\lambda$, the expression in
square brackets is none other than the sum of the work polarization $P_{W}^{n}$
over $N_{\mathrm{orb}}L$ single particle eigenstates of the one-dimensional
problem. Thus the average work polarization,
\begin{equation}
\overline{P}_{W}=\frac{1}{N_{\mathrm{orb}}L}P_{W}^{n}=\nu\hbar\Omega\label{eq:quant_work_polarization}
\end{equation}
is quantized as promised.

Let us now elaborate on how \eref{eq:quant_work_polarization} yields quantized
responses near the edge of a filled region as discussed in the main
text. Let us begin by considering the many-body Floquet eigenstate
$|\Psi(\lambda)\rangle$ obtained by filling all Floquet eigenstates
$\{|\psi_{n}(\lambda)\rangle\}$ that are located in a finite region
$S$ of the chain, of length $\ell$, as indicated in Fig.~1 from
the main text. For a site $x_{0}$ outside the filled region, the
force density $\rho_{x}^{F}$ is trivially zero for all values of
$\lambda$ and $t$, and hence the system can't absorb any energy
here. For a site $x_{0}$ in the bulk of the filled region, $\rho_{x}^{F}$
also vanishes when averaged over $\lambda$ and $t$: 
\begin{equation}
\bar{\rho}_{x}^{F}\equiv\frac{1}{2\pi T}\int_{0}^{T}dt\int_{0}^{2\pi}~d\lambda\langle\Psi(\lambda,t)~|\frac{\partial\rho_{x}^{E}(\lambda,t)}{\partial\lambda}|\Psi(\lambda,t)\rangle=0.\label{eq:BulkStateAbsorptionRate}
\end{equation}
To see how the above result follows, we insert the explicit form for
$\rho_{x}^{E}$ in second quantized notation, 
\begin{equation}
\rho_{x}^{E}(\lambda,t)=\frac{1}{2}\sum_{a}(H_{ax}(\lambda,t)c_{a}^{\dagger}c_{x}+H_{xa}(\lambda,t)c_{x}^{\dagger}c_{a}).
\end{equation}
Noting that that for sites $a,b$ in the bulk $\langle\Psi(\lambda,t)|c_{a}^{\dagger}c_{b}|\Psi(\lambda,t)\rangle=\delta_{ab}$
for all values of $t$ and $\lambda$, we find 
\begin{equation}
\bar{\rho}_{x_{0}}^{F}=\int_{0}^{2\pi}~d\lambda\frac{\partial H_{x_{0}x_{0}}(\lambda,t)}{\partial\lambda}=0.\label{eq:BulkStateAbsorptionRateProof}
\end{equation}
where the last equality follows from $H(2\pi,\lambda)=H(0,\lambda)$.
Finally, if the localization length is much less than the filled region
size, $\xi\ll\ell$, we may treat all the energy pumping as occurring
directly at the density edge and the result from the main text follows.

While the above arguments from dimensional reduction hold perfectly
in the case where states are fully localized and may be adiabatically
tracked upon varying $\lambda$, it is important to note that this
is indeed not the case for the models we consider except at the fine-tuned
point $W=\alpha=0$. We have indeed seen this in the numerical results
for finite disorder, as the lack of a well-defined adiabatic limit
gives rise to a smooth crossover from topological to non-topological,
rather than a sharp transition as in the two-dimensional case. Healing
the above arguments when this limit is not satisfied is a subtle issue
which we will address in a follow up work to appear shortly \cite{Upcoming_Long_Paper}.
In particular, we will show that a more careful derivation of the
quantized work polarization (or indeed the magnetization in the absence
of localization) gives rise to non-topological terms as well as the
topological contribution. We will argue that non-topological contributions
are suppressed exponentially as $e^{-L/\xi}$ in the presence of arbitrarily
weak localization, while the topological contribution is only slowly
destroyed in a system-size-independent manner as disorder strength
is increased.

\section{Quantized response of fine-tuned model}

Let us quickly see analytically that we achieve work quantization for the fine-tuned model with $J=J_\mathrm{tuned}$. This model has the nice property that for $t=nT/5$, the Floquet eigenstates are simply localized on each site. The force density is clear zero except during periods $2$ and $4$, as $h_{1,3,5}$ are independent of $\lambda$. Consider the bulk state indicated by the black arrow in Fig. 2a from the main text,  starting on the $A$ sublattice of site $x$. During step $2$, let us write the 2-site Hamiltonian connecting $|x,B\rangle$ to $|x+1,A\rangle$: $H_2 \to -J(\sigma^x \cos \lambda + \sigma^y \sin \lambda)$. Time evolving this effective spin-$1/2$ starting from the state $|\uparrow\rangle$, we see that $\langle \partial_\lambda H \rangle = -J \sin\left[\pi (t-T/5)/(T/5)\right]$, which averages to $2J/\pi$ over the period $T/5 < t < 2T/5$. This force is evenly split between sites $x$ and $x+1$. During step $4$, it has average force is similarly $-2J/\pi$, entirely on site $x$. Summing these up, along with zero responses during the other three steps, and integrating over $\lambda$, we see that the work polarization is
\begin{equation}
  P_W \equiv \int d\lambda \bar{P}_F = \left(\frac{1}{5}\right)\left(2\pi\right)\left(\frac{2J}{\pi}\right) \left[(x+1/2) - x\right] = \frac{\hbar \Omega}{2}.
\end{equation}
One may readily check that starting at site $|x,B\rangle$ gives the same result, such that the work polarization for a filled unit cell is the quantized value $\hbar \Omega$, as promised.

\end{widetext}

\end{document}